# Microcantilever-integrated photonic circuits for broadband laser beam scanning

Saeed Sharif Azadeh,[1,*] Jason C. C. Mak,[2] Hong Chen,[1] Xianshu Luo,[3] Fu-Der Chen,[1,2] Hongyao Chua,[3] Frank Weiss,[1] Christopher Alexiev,[2] Andrei Stalmashonak,[1] Youngho Jung,[1] John N. Straguzzi,[1] Guo-Qiang Lo,[3] Wesley D. Sacher,[1] & Joyce K. S. Poon[1,2,*]

[1] *Max Planck Institute of Microstructure Physics, Weinberg 2, 06120 Halle, Germany*

[2] *University of Toronto, Department of Electrical and Computer Engineering, 10 King's College Road, Ontario, M5S 3G4, Toronto, Canada*

[3] *Advanced Micro Foundry Pte. Ltd., 11 Science Park Road, Singapore Science Park II, 117685, Singapore*

*Emails: sazadeh@mpi-halle.mpg.de, joyce.poon@mpi-halle.mpg.de

## Abstract

Laser beam scanning is central to many applications, including displays, microscopy, three-dimensional mapping, and quantum information. Reducing the scanners to microchip form factors has spurred the development of very-large-scale photonic integrated circuits of optical phased arrays and focal plane switched arrays. An outstanding challenge remains to simultaneously achieve a compact footprint, broad wavelength operation, and low power consumption. Here, we introduce a laser beam scanner that meets these requirements. Using microcantilevers embedded with silicon nitride nanophotonic circuitry, we demonstrate broadband, one- and two-dimensional steering of light with wavelengths from 410 nm to 700 nm. The microcantilevers have ultracompact ~0.1 mm$^2$ areas, consume ~31 to 46 mW of power, are simple to control, and emit a single light beam. The microcantilevers are monolithically integrated in an active photonic platform on 200-mm silicon wafers. The microcantilever-integrated photonic circuits miniaturize and simplify light projectors to enable versatile, power-efficient, and broadband laser scanner microchips.

# Introduction

Optical beam steering is important for engineered light projection in numerous applications including displays [1], [2], microscopy [3]–[6], light detection and ranging (LiDAR) [7], [8], communications [9], [10], and ion/atom manipulation in quantum processors [11]–[13]. Most commonly, beam scanning is implemented with discrete components, such as galvo-scanners, micro-electromechanical systems (MEMS) mirrors, or acousto-optic deflectors [14]. In recent years, the demand to reduce the size of beam scanners into microchips for easier integration into products has motivated rapid advances in optical phased arrays (OPAs) and focal plane switch arrays (FPSAs) using silicon (Si) photonic integrated circuit (PIC) technology [15]–[17]. FPSAs illuminate discrete points, while OPAs offer continuous angular coverage. Not only can PIC beam scanners minimize size and power consumption, their co-integration with other components, such as photodetectors and lasers, onto a single chip can substantially simplify packaging and reduce costs compared to assemblies of light sources and beam deflectors [15], [18]–[27]. OPAs and FPSAs are very large-scale PICs consisting of arrays of hundreds to tens of thousands of grating coupler light emitters; and to date, they have predominantly been demonstrated in the infrared (IR) spectral region. To achieve two-dimensional (2D) beam steering in OPAs, wavelength sweep and phase-shifters are typically used in conjunction to reduce PIC complexity [23]–[27]. In the visible spectrum, OPAs and FPSAs are even more challenging to realize due to the lack of compact wavelength-tunable lasers and the lower efficiency of phase shifters and switches. Furthermore, the wavelength degree of freedom cannot be used for applications that require specific wavelengths, such as displays and the excitation of atomic transitions. Another major obstacle is that the half-wavelength pitch criterion for single-lobe emission in an OPA [15] is hard to satisfy in the visible spectrum without introducing significant inter-waveguide crosstalk or reducing the minimum feature size. Recent demonstrations of visible-light beam scanners have been based on OPAs [28]–[33], using an external super-continuum source coupled with a tunable filter [33] or a rotation stage to scan the beam in the second dimension while requiring high on-chip driving powers of 2 W [31]. These approaches are difficult to miniaturize into single chips in the foreseeable future.

      Here, for the first time to our knowledge, we demonstrate visible spectrum 1D and 2D beam scanner PICs that emit single beams without any sidelobes and at arbitrary wavelengths. The scanners consist of MEMS cantilevers with integrated silicon nitride (SiN) nanophotonic waveguides and grating couplers. The electrical drive power to cover the full scanning range of each axis was < 31 mW, about 2 orders of magnitude lower than previously reported visible light PIC beam scanners [31]. When resonantly driven, a scanning rate in the range of 10-100 kHz was achieved dependent on the cantilever length. In contrast to the recent Si waveguide MEMS phase-shifters and beam scanners [16], [31], [34],

our approach does not require waveguiding in Si or electrically conductive Si. Our cantilevers are agnostic to the waveguide core material; hence, they apply to SiN waveguides, which are optically transparent at visible wavelengths. Using standard fabrication processes in Si photonics foundries, our cantilever devices were monolithically integrated within a foundry-manufactured visible spectrum PIC platform on 200-mm Si (Figures 1a,b), in which other components including low-loss wideband edge couplers, high quantum efficiency waveguide photodetectors, low crosstalk junctions, and efficient thermo-optic phase shifters have been reported [35]–[38]. Due to the mechanical nature of the actuation, identical steering ranges were achieved for wavelengths between 410 to 700 nm. Because our approach does not require a large array of light emitters, excluding the laser source, our PICs possess the smallest footprint amongst all chip-scale beam scanners to date of 0.14 mm × 1.1 mm for 2D scanning. These versatile MEMS cantilevers can also be easily implemented in generic silicon-on-insulator (SOI) photonic platforms with heaters, deep trenches, and an undercut step, opening new directions for compact beam-scanning PICs.

## Results

### Operation principle and architecture

We realized two types of beam scanner designs: (1) a rectilinear MEMS cantilever (Figures. 1c, e), which steered the output beam only in the longitudinal direction, and (2) an L-shaped singly clamped cantilever (Figures. 1d, f) capable of beam steering in both the longitudinal and transverse directions via two control voltages. In both cases, light was guided in a SiN waveguide embedded in the cantilever and terminated with an output grating coupler at the distal end. The grating coupler was 10 μm wide and 25 μm long, consisting of fully etched 150 nm thick SiN teeth with a period of 440 nm. It had an average loss of 5.2 dB at wavelengths between 410 and 700 nm (see Supplementary Section 1 for details on the grating couplers). The simulated grating coupler emission full-width at half-maximum (FWHM) beam widths were 0.78° longitudinally and 2.2° transversely, and the measured FWHM widths were 1.4° and 3.1°, respectively. Figure 1a shows the PIC platform cross-section with the cantilever delineated. The suspended structure was formed by a deep trench $SiO_2$ etch followed by an isotropic Si undercut. To actuate the cantilever, we created an electro-thermal bimorph using a 2 μm-thick aluminum (Al) layer atop a 2.5 μm thick $SiO_2$ layer (see Supplementary Section 2 for the layer thickness design). In contrast to [38] where the lower level of the Si mesa was undercut, here, the higher Si level was undercut to achieve the $SiO_2$ thickness in the cantilever. Embedded resistive titanium nitride (TiN) strips heated the cantilever with applied voltages as shown in the circuits in Supplementary Section 3 Fig. S3. Due to the greater thermal expansion of Al compared to $SiO_2$, the cantilever bends downward with increasing temperature.

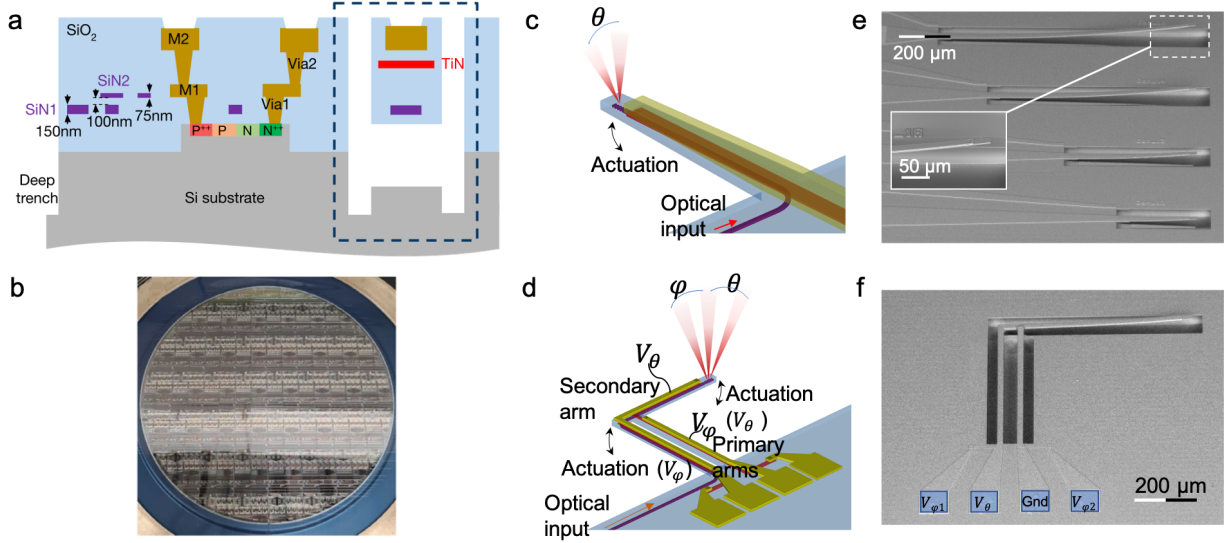

**Fig. 1: Overview of the microcantilevers. a** Cross-sectional schematic of the integrated photonic platform containing the microcantilevers. The cantilever cross-section is delineated in the dashed box. **b** Photograph of the fabricated 200mm diameter wafer. **c** Schematic of the rectilinear cantilever for 1D beam scanning. A SiN waveguide with an output coupling coupler and a TiN heater are embedded in the cantilever. **d** Schematic of the L-shaped cantilever for 2D beam scanning in the longitudinal (θ) and transverse (φ) directions. Scanning electron micrographs (SEMs) of **e** rectilinear cantilevers (with 300, 500, 800 and 1000 μm lengths) and **f** an L-shaped cantilever. Due to film stress, the cantilevers bend upwards in the absence of applied electrical power.

## *Rectilinear cantilevers*

Figure 1c illustrates the schematic of the rectilinear cantilever for 1D scanning. The width of the cantilever tapered from 30 μm at the proximal end to 15 μm at the distal end. This shape strikes a balance between robustness and thermal efficiency. A wider cantilever base is less likely to crack and has better performance under high stress, while a narrower cantilever width reduces the volume to be heated, and thus higher temperatures could be reached for the same applied electrical power, $P_e$. We designed rectilinear cantilevers with four different lengths ($L_{can}$) of 300, 500, 800, and 1000 μm, with resistances of the TiN ($R_{TiN}$) heaters 430, 480, 550, and 680 Ω respectively. The choice of cantilever length is a trade-off between the steering range and actuation time constant. As will be shown, while the longest cantilever achieved the largest steering range, the shortest devices were faster due to a lower thermal time constant.

In the absence of applied electrical power, $P_e$ = 0 mW, the cantilever bent upwards due to the initial stress between the metal and oxide, which were deposited at different temperatures. A scanning electron micrograph (SEM) of the rectilinear cantilevers captured at a 45° tilt shows this expected initial upwards bending (Figure 1e). The simulated angular steering range as a function of $P_e$ at $L_{can}$ = 500 μm is shown in Figure 2a. The top and bottom insets of Figure 2a illustrate the downward displacement of the cantilever tip under applied power. The tuning efficiency ($d\Delta\theta/dP_e$) is reduced at $P_e$ > 14 mW as the cantilever tip contacts the Si substrate at the bottom of the undercut, modeled to be 25 μm deep, yet continues to bow (see Supplementary Section 3 for details). The initial strain was calculated assuming the deposition temperature of Al to be 88°C to match the experiment. The simulations predict an angular scan range of 17.6° and 29.5° respectively for the 500 and 1000 μm long cantilevers at $P_e$ = 30 mW.

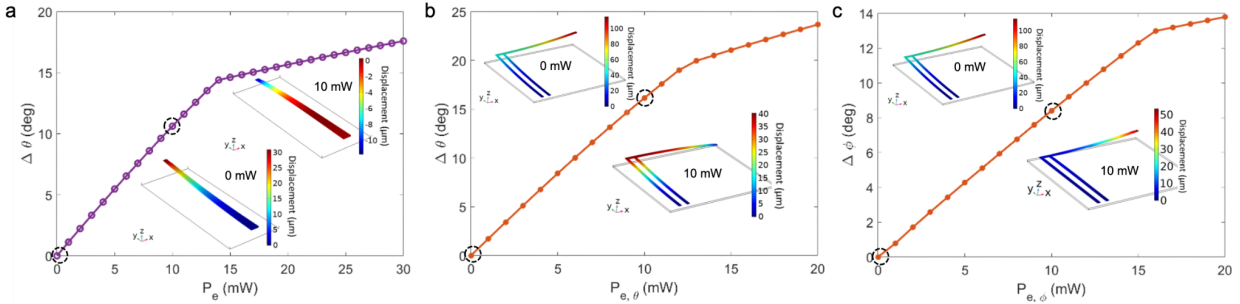

**Fig. 2: Simulated beam steering of the rectilinear and L-shaped microcantilevers. a** Simulated angular scan range of the 500 μm long rectilinear cantilever vs. applied electrical power ($P_e$) to the TiN layer; insets show the simulated shape of the cantilever under 0 mW (bottom) and 10 mW (top) of applied power. **b** Simulated angular scan ranges of the L-shaped cantilever along φ-axis and **c** θ-axis. The lengths of the primary and secondary arms are 500 and 600 μm, respectively. The insets show the calculated displacement of the L-shaped cantilever under 10 mW (bottom) and 0 mW (top) electrical power applied to the **b** primary arm and **c** secondary arm.

### *L-Shaped Cantilevers*

Figure 1d shows a schematic of the L-shaped cantilever for 2D beam scanning. The design has two primary arms that tilt the grating coupler in the transverse direction under a control voltage $V_φ$ and a secondary arm which tilts the beam in the longitudinal direction with a control voltage $V_θ$. An SEM image of the device is shown in Figure 1f. The primary and secondary arms were 500 and 600 μm long, respectively, with a constant width of 20 μm. The SiN waveguide in the cantilever had a 40 μm bend radius to connect the input to the grating coupler at the cantilever tip. Figures 2b and 2c show the simulated beam steering along the φ-axis with respect to applied power to the primary arms ($P_{e,φ}$), and along θ-axis with respect to electrical power applied to the secondary arm ($P_{e,θ}$), respectively. Thermal crosstalk between the primary and secondary arms causes a slight θ-axis tilt under $P_{e,φ}$, and vice versa (see Supplementary Section 3), but is sufficiently small to allow for independent control of the beam direction along the two angular axes. Again, the reduction in the angular tuning efficiency is due to the tip of the cantilever coming into contact with the substrate. The simulations predict a maximum beam steering of 23.7° along the longitudinal direction under $P_{e,θ}$ = 20 mW, and 13.8° under $P_{e,φ}$ = 20 mW.

## Measurements

### *1D beam-steering with the rectilinear cantilevers*

The far-field radiation pattern was captured using the setup described in the Methods Section and illustrated in Figure 3a, where the input laser light is coupled from a multi-wavelength laser source via a single-mode fiber onto the chip. Unless otherwise stated, the input polarization was set to transverse-electric (TE) mode to maximize the optical transmission of the grating coupler. The recorded far-field image of the grating output shows a divergence angle of 1.4° in the longitudinal direction (θ) and 3.1° in the transverse direction (φ) at λ = 488 nm. As the beam steering range is independent of the laser wavelength due to the mechanical nature of the actuation, λ = 488 nm was chosen as a representative wavelength for measurements unless otherwise stated. Figure 3b visualizes the extent of the steering range of the four rectilinear cantilevers, by capturing an overlay of the far-field images at applied powers of 0 mW (right beams) and 20 mW (left beams).

The measured beam steering angles as a function of the applied electrical power are shown in Figure 3c. We measured maximum beam scanning ranges of 11°, 17.6°, 22.6° and 30.1° with 30 mW applied electrical power, respectively, for 300, 500, 800, and 1000 μm long cantilevers, in good agreement with the simulated values (dashed lines). This corresponds to ~ 8, 12, 16, and 21 resolvable points at λ = 488 nm, respectively for the shortest to the longest cantilevers. The power efficiency in terms of $\frac{d\Delta\theta}{dP_e}$ of the 1000 μm cantilever was 1°/mW and was lower for shorter cantilevers due to the heat sinking effect of the metal contacts, which reduced the effective temperature in the proximal end of the cantilever. As predicted by the simulations and attributed to the cantilever coming in contact with the substrate (Figure 2, Supplementary Section 3), a reduction in the angular tuning efficiency was observed.

The non-resonant scan rate was mainly limited by their thermal time constant. We measured the time response of the devices by applying a 10 mW pulsed signal with various duty cycles and recording the far-field pattern (see Methods Section and Supplementary Section 4). Figures 3d and 3e, respectively, show the measured 10%-to-90% rise and fall time responses of the 300 μm long cantilever (see Figures S4 and S5 for the time response of the other beam scanners). We measured an average response time of 1.2, 2.6, 4.1 and 4.7 ms, respectively, for the shortest to the longest cantilevers. These response times are comparable to the fastest liquid crystal switches [39]. However, many beam steering applications (including displays, 3D sensing, and microscopy) require only periodic scanning of light beams in at least one direction and not aperiodic switching. Thus, to reach higher scanning rates, we drove the cantilevers at their resonance frequencies, beyond the electro-thermal time-constant limit (Figure 3f). The resonance frequencies of the devices were 5.7, 11.8, 24.8, and 77.4 kHz respectively for 1000, 800, 500, and 300 μm long cantilevers. At these frequencies, maximum beam scan ranges of 12°, 10.4°, 11.3°, and 9.8° were achieved under 20 mW of applied electrical power, respectively.

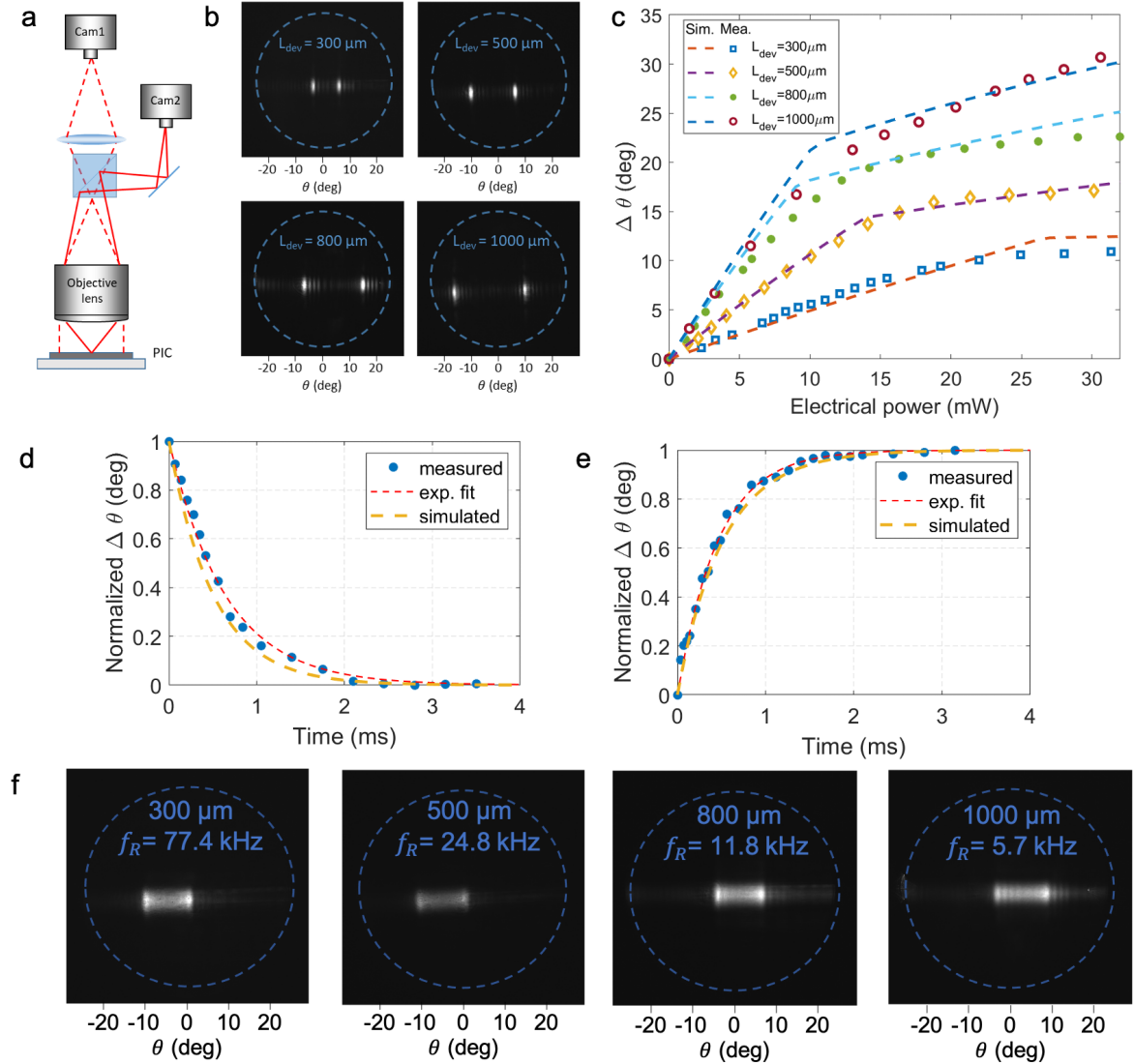

**Fig. 3. Rectilinear cantilever: Experimental setup, far-field patterns, DC, and time-dependent response. a** Schematic of the imaging setup using a regular lens L for creating a Fourier image, and a beam splitter BS for simultaneous capturing of the near- and far-field patterns, with solid lines and dashed lines showing, respectively, the near-field and far-field trajectories. **b** Measured far-field patterns of the grating output at 0 mW (right beams) and 20 mW (left beams) for the four cantilever lengths. The two outputs are overlaid on the same image by applying a 2 Hz step function and capturing the grating far-field output over a 1-second exposure. **c** Measured (dots) and simulated (dashed lines) beam steering versus applied DC electrical power to the rectilinear cantilevers. **d** Fall time and **e** rise time of the 300 μm long cantilevers. **f** Far-field images of the rectilinear cantilevers captured at their respective resonance frequencies.

## *2D beam steering with the L-shaped cantilevers*

The measured angular scan ranges of the L-shaped cantilever are shown in Figures 4a and 4b, respectively, in the longitudinal and transverse directions. The lengths of the primary and secondary arms were 500 and 600 μm, respectively. An angular steering range of 24.0° and 12.2° was achieved with a maximum applied electrical power of 23 mW, respectively, for the θ-axis and φ-axis. To illustrate the 2D angular range of the steering in Fourier space, we applied two different signals to each set of arms. A 120 Hz sinusoidal electrical signal with a peak-to-peak voltage of 4 V and a DC offset of 2 V was applied to the secondary arm, while a 30 Hz electrical signal with a peak-to-peak voltage of 3 V and DC offset of 1.5 V was applied to the primary arms. The corresponding far-field image of the output beam captured over a 33 ms exposure time is shown in Figure 4c, covering a range of ~ 24°×12° in Fourier space.

The 10-90% rise time of the primary and the secondary arms were 4.3 and 4.7 ms, but faster beam steering is possible on resonance. Figure 4d shows the simulated first (left) and second (right) resonance modes of the L-shaped cantilevers, which were at 6.9 kHz and 14.3 kHz, respectively. At the first resonance, both the secondary and the primary arms were in phase, simultaneously moving upward (or downward) thus moving the Fourier image of the far-field beam in the negative (or positive) directions of the θ- and φ-axes. Experimentally, the first resonance frequency was found to be 7.6 kHz, with its far-field radiation pattern shown in Figure 4e. At the second resonance frequency, measured to be 17.4 kHz, the displacement of the primary and secondary arms had a π-phase difference, resulting in the far-field pattern in Figure 4f. To excite the resonances, a pulsed voltage with an average power of 10 mW was applied to the primary arms of the L-shaped cantilever. These resonances can be excited in linear superposition, as shown in Figure 4g.

As a proof-of-concept demonstration, we used the cantilever to project a 2D image of our department name "NINT" (Figure 4h). The applied voltages to the primary and secondary arms were controlled by a computer to steer the beam in the Fourier space in the desired directions. The images were generated without controlling the laser output power, and the pattern formation relied solely on the actuation of the beam. The refresh rate of the image was set to 30 Hz over a 33 ms exposure time. The experimental results of 1D and 2D beam scanners are summarized in Table S1.

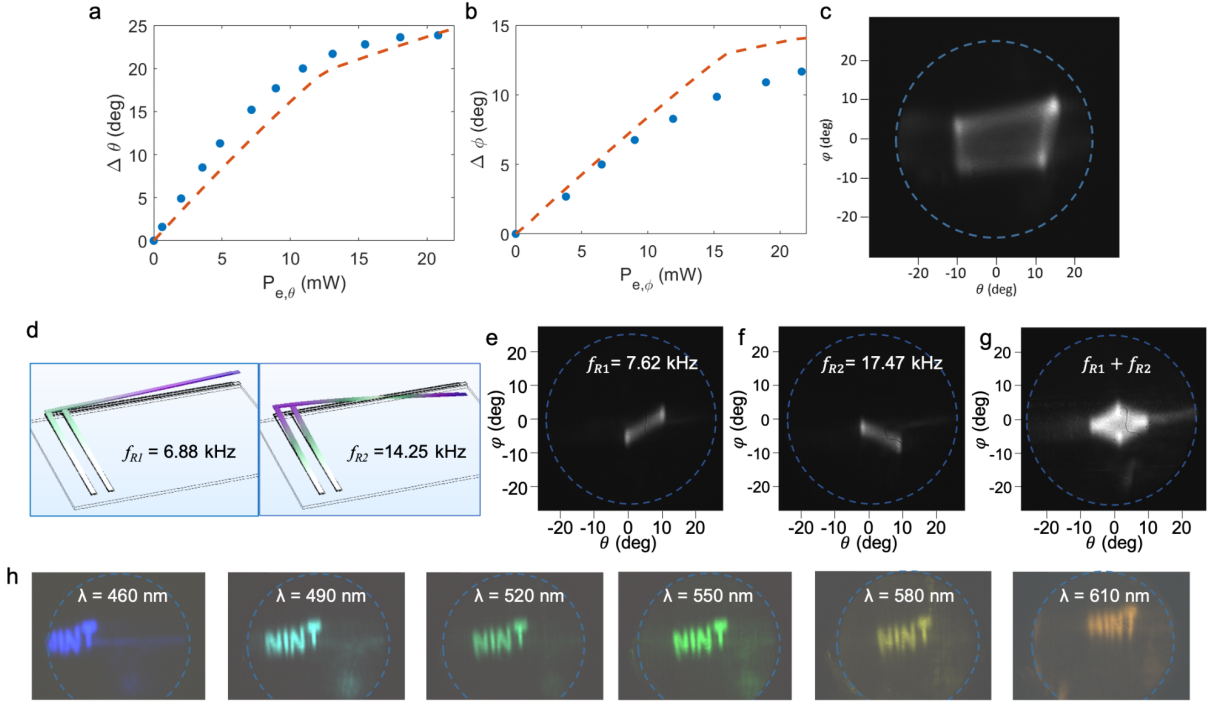

**Fig. 4. L-shaped cantilever characterization. a (b)** Measured steering of the output beam in longitudinal (transverse) direction vs. the electrical power applied to the secondary (primary) arm. **c** Recorded far field pattern of the L-shaped cantilever output under a drive voltage $V_\theta = 2(2\pi f_1 t) + 1] V$, where $f_1$ = 120Hz, applied to the secondary arm, and $V_\phi = 1.5(2\pi f_2 t) + 1] V$, where $f_2$= 30Hz applied to the primary arms. **d** Simulated resonance modes of the L-shaped cantilevers at the first and second resonance frequencies. Far-field images of the device are recorded at the first (**e**) and the second (**f**) resonance frequencies, and superposition (**g**) of the two resonances. **h** Images produced by the L-shaped cantilever scanning our department name ("NINT"), demonstrating the potential for image projection.

Finally, we characterized our beam scanner in cryogenic conditions. Another L-shaped cantilever device was tested in a cryostat at temperatures as low as 10 K (see Methods and Figure S7 for details on the cryogenic unit). By reducing the temperature, the cantilever was further bent upwards due to the higher expansion coefficient of Al compared to SiO$_2$. Therefore, the initial angle of the output beam was increased in both θ and φ directions, as shown in Figure 5a, where the simulated initial angle (lines) and the measured values of the tilt (dots) are plotted. The simulated scan range of the L-shaped cantilever versus the applied electrical power in θ and φ directions are respectively shown in Figures 5b, c. The dashed line shows the simulated scan range at room temperature for comparison. Due to the higher initial deflections at 10 K compared to room temperature, the beam can deflect a larger angle without contacting the Si substrate; thus, no change in the angular tuning efficiency was observed at up to 40 mW of applied electrical power. The experimental results (yellow dots in Figures 5b, c) are in good agreement with the simulated tilt of the cantilever. Due to the limited field-of-view of the cryogenic setup (i.e., the emitted beam incident on the sides of the cryogenic chamber), the experimental observations were limited to angles less than 20°. The cantilever was driven at a resonance frequency of 16.76 kHz for ~7 million cycles at 10 K with no observable degradation. Thermal simulations show that the heat is localized to the cantilever device (Supplementary Section 6). The calculated half-decay length of the temperature, defined as the distance where the difference between the local temperature and the cryostat setting is 50% of the maximum, is 20 μm for 26 mW of applied power, and the temperature on the PIC drops to below 11 K within 600 μm distance from the edge of the cantilever.

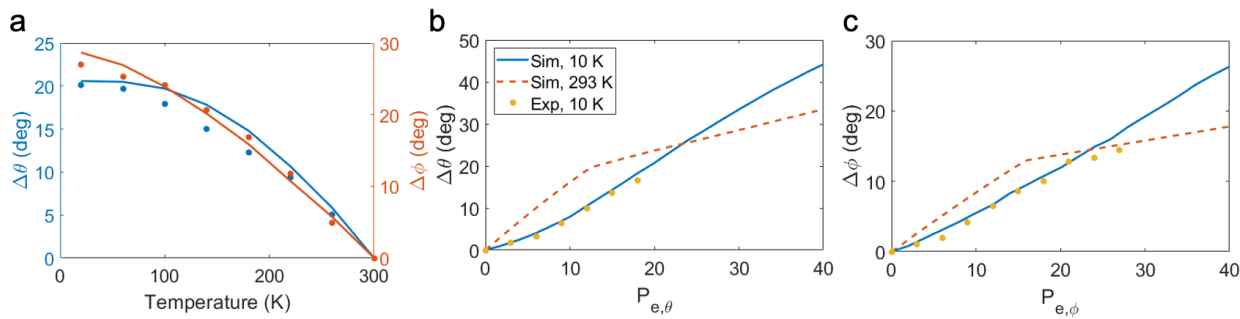

**Fig. 5: Cryogenic characterization of the L-shaped microcantilever. a** Simulated (solid lines) and measured (dots) values of the initial angle shift vs. the ambient temperature in the longitudinal (left axis) and transversal (right) directions, without applying electrical power. **b (c)** Simulated and measured steering of the output beam in longitudinal (transverse) directions.

## Discussion

These reported ultracompact, power efficient, monolithically integrated microcantilevers are, to the best of our knowledge, the first side-lobe-free 2D beam scanners for the visible spectrum and the beam scanners with the widest operation wavelength bandwidth spanning >300 THz. Our devices achieved a 2D scan range of 24°×12° with arm lengths of ~500 μm with only 46 mW of drive power. The beam scanning was achieved without wavelength-tuning or using phase shifters at scan rates of tens of kHz on resonance. The scanning range and power emission of the devices were unchanged after one billion scan cycles on resonance in ambient conditions.

In comparison with the prior art, our device offers unique advantages (see Supplementary Section 7 for a comparison chart). Our approach is distinct from the photonic-MEMS phase-shifters using Si waveguides [16], [34] since we do not require highly doped Si. It also achieves single-lobe output beam steering without wavelength tuning, which has not yet been possible with visible spectrum OPAs [28], [31], [40]. The power consumption of the cantilever beam scanners was approximately 2 orders of magnitude lower than visible-wavelength OPAs, which required ~2 W [31]. Our beam scanners are also distinct from previous MEMS-tunable grating couplers for spectral tuning and fiber-to-chip coupling in the infrared [41]–[45]. First, we have achieved a larger 1D scan range (30° in 1D vs. 5.6° in [44]) and 2D scanning for the first time [45]. Second, our design displaces the entire grating emitter rather than tuning the grating period and apodization, so the emission profile minimally deteriorates during cantilever actuation. For example, in [44], due to the wide angular beam width, the number of resolvable points is only about 0.62. In comparison, the number of resolvable spots of the L-shaped cantilever here is about 66, limited by the divergence angle of the grating emission (1.4° in θ-direction and 3.1° in φ-direction). Lastly, the microcantilevers dramatically reduce the complexity of the drive circuitry for 2D beam scanning compared to large-scale PIC approaches of OPAs and FPSAs – only a single device needs to be controlled with 2 drive signals (for 2D beam scanning) and resonant scanning enables ~100 kHz scan rates. This simplification will reduce the power consumption of the drivers, calibration, control, and stabilization of a large number of array elements.

The cantilever beam scanners can be extended in several ways. The achieved scan range was limited by the substrate, so it can be increased by making the undercut deeper. Since the initial deflections are determined by the beam lengths, multiple beam scanners can be used together to expand the scan range. To increase the number of resolvable spots, the divergence angle of the output beam along the propagation direction could be significantly decreased by >10× using weaker and longer gratings (Supplementary Section 5). To reduce power consumption, electro-static or piezoelectric actuators can be incorporated instead of the electro-thermal bimorph design [46]. Beyond the demonstrated geometry, other planar-compliant mechanisms, as well as other types of light emitters, such as edge couplers, sub-wavelength waveguides, OPAs, and metasurfaces, can be used. The deformation along the cantilever may also be exploited as a tuning method for or a sensor using embedded nanophotonic devices.

In summary, the microcantilever-integrated photonic circuits demonstrated here open exciting avenues for photonic beamforming. The approach decouples the design of the scan range from the light emitter. The cantilevers are simple to control, can be incorporated in any photonics platform possessing an undercut etch, and can be placed anywhere within a chip. Microcantilever-integrated photonic integrated circuits may enable ultracompact and power-efficient solutions to transform augmented reality displays, microscopy, quantum information processors, and 3D mapping technologies.

## Methods

### Numerical simulations

Electro-thermomechanical simulations of the MEMS structures were performed using finite element method (FEM) in COMSOL Multiphysics to find the resonance modes of the cantilevers, as well as their displacements with respect to the applied voltage, and their time response. The Young's modulus of SiN, TiN, Al, and $SiO_2$ were assumed to be 250, 500, 70, and 73 GPa and their Poisson ratio was set to 0.23, 0.25, 0.33, and 0.17, respectively. For thermal simulations, the Si substrate and electrical pads were set to a constant temperature of 293 K. The thermal expansion coefficients of Al and $SiO_2$ were set to $23\times10^{-6}$ and $5.5\times10^{-7}$ 1/K, respectively, with thermal conductivities of 238 and 1.4 W/mK. Optical simulations of the grating couplers were carried out using the 3D finite difference time domain (FDTD) method in Lumerical software. The refractive indices of SiN and $SiO_2$ were assumed to be 1.81 and 1.46 at $\lambda = 532$ nm.

## Device fabrication

The devices were fabricated on 200-mm diameter Si wafers at Advanced Micro Foundry (AMF) as part of our visible-light photonic integrated circuit platform. The fabrication process included steps to implement other devices in this platform. It started with ion implantation and partial etching of the Si substrate to define the photodetectors [36]. Next, a $SiO_2$ layer as the bottom cladding of the waveguides was formed using PECVD. Then a SiN layer with the targeted thickness of 150 nm was deposited atop the oxide layer in a PECVD process. The SiN waveguides were then defined by 193nm deep ultraviolet (DUV) lithography followed by a reactive ion etching (RIE) step. Additional $SiO_2$ and SiN deposition and patterning steps were performed to define a second 75 nm thick SiN waveguide layer to form low-loss bi-layer edge couplers [35]. The layers were planarized using chemical mechanical polishing. Next, a TiN layer was deposited and patterned to be used as a heater, followed by two Al layers and oxide openings for bond pads. The top Al layer thickness was 2 μm to enhance the strain and thus the initial displacement of the cantilevers. Finally, to suspend the MEMS structures and also to form the $SiO_2$ bridges in our thermal phase-shifters [38], a deep trench was formed followed by undercut etching of the Si.

## Room temperature measurement setup

Figure 2a shows the experimental setup for device characterization. The setup captured the emission pattern in real and Fourier space imaging modes. The far-field output was collected by a high numerical aperture objective lens (with 20× magnification, NA = 0.42, and effective focal length = 10 mm) to project the far-field radiation pattern into the Fourier plane, where it was captured by a CMOS camera. We utilized an uncollimated white light source to visualize the sample surface (not shown in Fig. 2a). For simultaneous visualization of the near-field and facilitating the alignment procedure, a beam splitter diverted half of the radiated beam to a second CMOS camera. Light from a multiwavelength laser source (Coherent OBIS Galaxy) was edge-coupled to the chip through a single-mode fiber (Nufern S405-XP) with an inline polarization controller. The polarization was set to transverse electric (TE) mode to maximize the optical transmission of the grating coupler.

## Time response measurements

To measure the temporal response of the cantilevers, we coupled light into each device and recorded the far-field radiation under an applied periodic electrical pulse with a peak power of 10 mW and varying duty cycles. In the case of a 50% duty cycle, provided that the period of the square pulse (T) was much longer than the rise/fall time of the cantilever ($t_r$), the maximum displacement of the far-field beam would be equal to the results for a DC voltage (the first far-field image in Fig. S4a). By reducing the duty cycle to a level below the rise time of the device, the emitted beam trajectory became shorter, allowing us to determine the transient response of the device. Measurements of the far-field trajectories are shown in Supplementary Section 4.

Cryogenic measurement setup

The cryostat was taken to a vacuum at a pressure of $< 10^{-4}$ mbar and the temperature was reduced by liquid Helium (He) cooling. The temperature of the cold head was controlled using a 100 W built-in heater connected to an automatic PID controller, which could also set the He flow using a magnetic valve. To establish electrical connectivity, the PIC was mounted on a custom printed circuit board (PCB) using a thermally conductive epoxy (Loctite 84-1LMIT1) and then wire bonded to a PCB (Fig. S7). The wires as well as the optical fiber were routed inside the chamber via electrical and optical high-vacuum fit-through adapters. The optical fiber was attached on top of the PCB using a transparent optical adhesive (DYMAX OP-4-20632) and cured with UV light while being actively aligned to the input edge coupler.

## Data Availability

The data that support the findings of this study are available from the authors on reasonable request.

## Acknowledgements

The authors thank Dr. Holger Meyerheim for lending the cryostat.

# Supplementary Information

## 1. Simulation and characterization of the grating couplers

The same 25 μm long grating coupler design was used in all devices presented in the manuscript. The grating coupler was fabricated by fully etching the 150 nm thick SiN waveguide layer. The width of the grating coupler was 10 μm. A 200 μm long adiabatic taper with an initial width of 380 nm (waveguide width) and a final width of 10 μm (grating coupler width) ensured only the fundamental TE mode was launched into the grating coupler. The lengths of the teeth and grooves were 220 nm. Figure S1a shows the simulated far-field pattern of the grating output beam at $\lambda = 488$ nm. The simulations were performed using three-dimensional FDTD in Lumerical. The far-field pattern along the longitudinal axis (cutline A in Fig. S1a) is plotted in Fig. S1b, and shows a full-width-half-maximum (FWHM) of 0.78° for the output beam. Similarly, the far-field pattern along the transverse direction (cutline B in Fig. S1a) predicts an FWHM of 2.2° as shown in Fig. S1c. As mentioned in the manuscript, the measured far-field image of the grating coupler had an FWHM of 1.4° and 3.1°, respectively in longitudinal and transverse directions. The difference between the simulation and experimental results can be attributed to fabrication variation, as the teeth width of 220 nm was close to the minimum feature size.

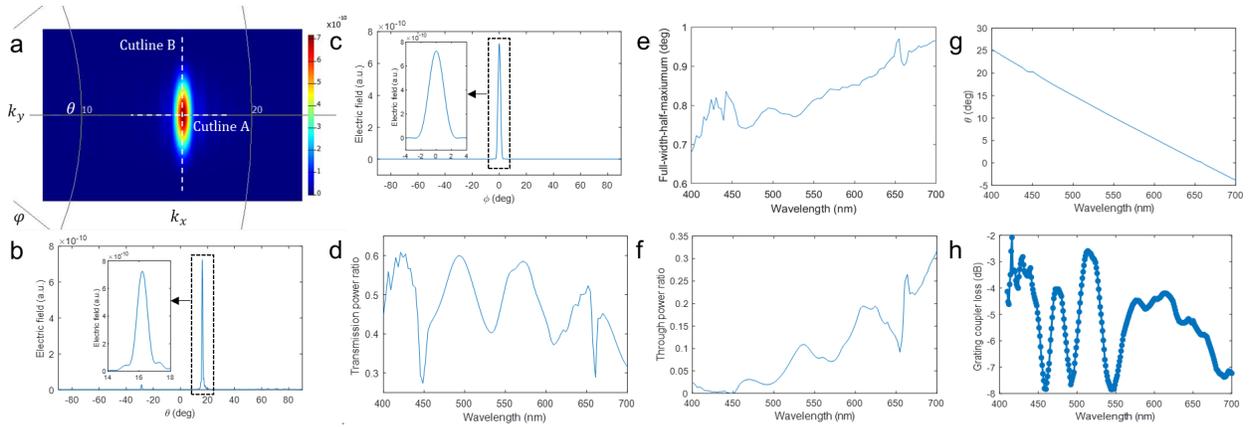

**Fig. S1: Grating couplers performance. a** Simulated far-field pattern of the grating coupler output at λ=488nm. **b** one-dimensional representation of the output beam along Cutline A ($k_x$). **c** One-dimensional representation of the output beam along Cutline B ($k_y$). **d** Simulated off-chip transmitted power of the grating coupler normalized to input power. **e** Simulated full-width-half-maximum (FWHM) of the grating coupler output. **f** Simulated on-chip optical power remaining at the end of the grating coupler normalized to the input power. **g** Simulated output beam angle of the grating coupler. **h** Measured optical loss of the grating couplers.

Figure S1d shows the simulated optical power transmission of the grating coupler normalized to the launch power in the wavelength range between 400 and 700 nm. Peaks and valleys in the transmission spectrum of the grating are expected due to the interference between the emitted beam and the back-reflected beam from the substrate. Based on our simulations, the depth of the fluctuations can be minimized by reducing the distance between the grating coupler and the substrate. The simulated FWHM spectrum of the output beam along the propagation direction is shown in Fig. S1e, and is < 1° in the visible spectrum. The ratio of the power remaining at the end of the 25 µm long grating coupler is negligible at λ = 488 nm, as shown in Fig. S1f. The simulated wavelength dependence of the output beam angle is shown Fig. S1g, which is in good agreement with the measured value of ~ 0.1°/nm. Finally, the measured transmission spectrum of the grating coupler is shown in Fig. S1h, which is slightly different from the simulated transmission spectrum (Fig. S1d) in terms of optical loss at a given wavelength. The discrepancy can be due to the high sensitivity of the optical transmission spectrum to the thickness of the oxide. For this measurement, we used a supercontinuum laser source coupled to an external tunable optical filter with a 1 nm optical bandwidth. At each wavelength, the polarization of the launch beam is set to TE in order to maximize the transmission of the grating coupler.

## 2. Selection of layer thicknesses for the Al-SiO$_2$ bimorph cantilever

Generally, bending can be induced in a two-material laminate by a difference in their thermal expansion coefficients. Bimaterial (or bimorph) cantilevers have been extensively studied in the context of bimetallic thermostats, such as the classic result of Timoshenko [1], and more recently as MEMS actuators that use metal and Si [2] or SiO$_2$ [3]–[6]. Bimaterial cantilevers can be formed in silicon (Si) photonic platforms by repurposing existing process features. Si photonic platforms typically include a thick top Al layer for electrical routing and bond pads that is on top of the SiO$_2$ cladding. As well, undercut etches remove a portion of the substrate to suspend the oxide and top metal layers. Undercut trenches are typically used for improving thermal isolation for thermo-optic phase shifters [7] and suspended edge couplers [8]. Bimorph cantilevers can be formed by an undercut under a region with a top metal layer. Optical circuits can be incorporated inside the cantilevers.

The deflection, $d$, of an Al-SiO$_2$ bimorph cantilever with the same base and tip width follows the equation [9]:

$$d = \frac{cx(1+x)^2}{c^2x^4+4cx^3+6cx^2+4cx+1} \frac{3\Delta\alpha\Delta T L^2}{t_{Al}+t_{SiO2}},$$

where $x = \frac{t_{Al}}{t_{SiO2}}$ is the ratio of the thicknesses of Al and SiO$_2$, $c = \frac{E_{Al}}{E_{SiO2}}$ is the ratio of the Young's modulus, $\Delta\alpha = \alpha_{Al} - \alpha_{SiO2}$ is the difference of thermal expansion coefficients, $\Delta T$ is the temperature change from initial, and $L$ is the cantilever length. The deflection is maximized when the thickness ratio of the metal to elastic material is of the ratio [9]

$$\frac{t_{Al}}{t_{SiO2}} = \sqrt{\frac{E_{SiO2}}{E_{Al}}}.$$

For values assumed in our numerical simulations, $E_{Al} = 70\ GPa$, $E_{SiO2} = 73\ GPa$, $\frac{t_{Al}}{t_{SiO2}} = 1.02$. In our platform, due to compromise with other features, this ratio was $\frac{t_{Al}}{t_{SiO2}} = 0.8$. Based on the deflection equation, this nonoptimal ratio is expected to decrease the maximum deflection by 1.5%. The deflection in the measured device will also be limited by the yield strength at the highest temperature reached. Alloys of Al which are CMOS compatible can be investigated to further improve reliability and range of the deflection. Beyond cantilevers, advanced bimorph actuator geometries and structures such as those in [6] can also be combined with integrated photonics in the future.

## 3. Thermal and deflection simulations of the cantilevers

### 3.1 Rectilinear cantilevers

Figure S2a shows the simulated temperature along the 500 μm long rectilinear cantilever under 10 mW of applied electrical power. The average temperature of the cantilever for a given electrical power is inversely proportional to the cantilever volume. Hence, reducing the thickness and width of the device results in a higher power efficiency. However, a narrower width worsens the mechanical robustness of the cantilever, especially for longer devices. Moreover, the thickness of the $SiO_2$ must be determined to suit the functionalities of other devices on the platform such as bilayer directional couplers. Shorter rectilinear cantilevers have a lower power efficiency due to the heat-sinking provided by the metal lines that form the electrical connections on the clamped side of the cantilever. Fig. S2b shows the simulated temperature along the 500 μm long cantilever is lower near the clamped end. Fig. S2c plots the displacement along the cantilever, and for an applied electrical power > 15 mW, its distal end comes into contact with the substrate, and the cantilever starts to bow upward near its center. This effect leads to the change in the slope of the angular tuning curves in Fig. 3c of the main manuscript.

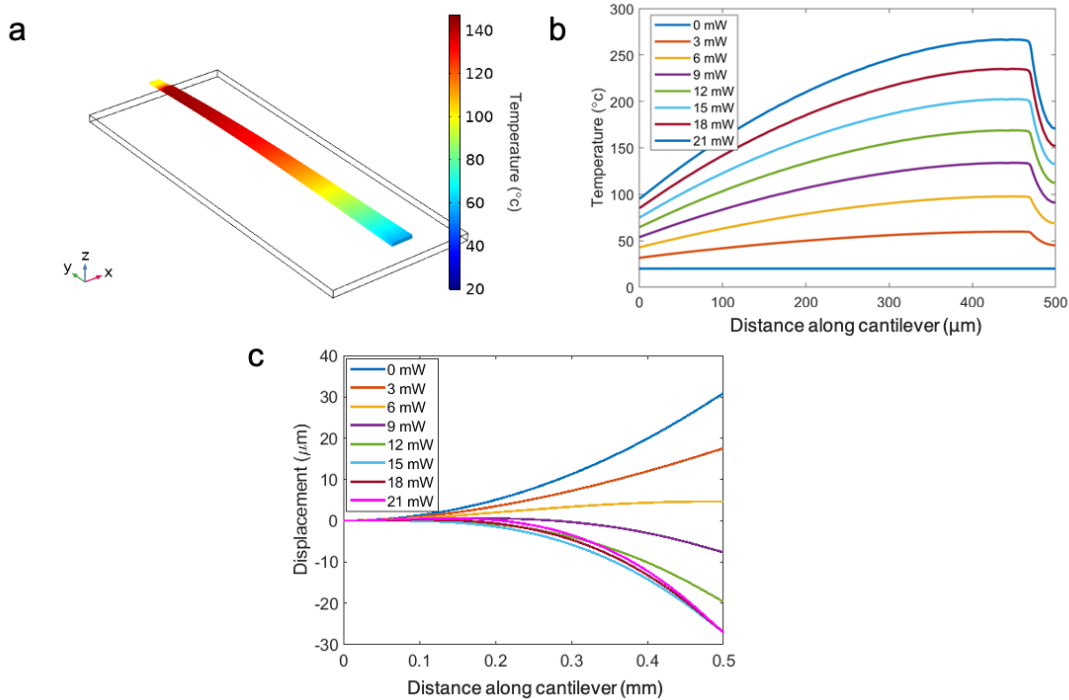

**Fig. S2: Electro-thermal simulation of a 500 µm long rectilinear cantilever. a** Simulated temperature distribution under applied electrical power of 10 mW. **b** Calculated temperature and **c** displacement along the cantilever for different electrical powers.

3.2 L-shaped cantilevers

In an ideal 2D beam steering system, the steering of the output beam in the longitudinal direction (θ) should be fully independent of the control voltage of the transverse direction ($V_\varphi$), and vice versa. For the L-shaped cantilever, this means that the electrical power applied to the secondary arm ($P_\theta$) should not tilt the grating coupler in the transverse direction. However, as shown in Fig. S3a, the temperature of the secondary arm under 10 mW electrical power has a tail that extends to the primary arm on the left side, causing it to slightly bend down and resulting in a parasitic angular tilt (Δφ) in the transverse direction. The resulting Δφ is calculated and shown in Fig. S3b. Nevertheless, the angular tilt in the desired direction (as shown in Fig. 4b, c) dominates the tilt due to the thermal crosstalk between the arms, and thus the beam can still scan over a wide range (Fig. 4d) in Fourier space. As shown in Fig. S3a, the thermal crosstalk between the secondary arm and the left primary arm is higher compared to the primary arm on the right side, due to the metal connection between these two arms electrically connect the secondary arm. Similarly, applying electrical power to the primary arms results in a slight temperature increase in the secondary arm (Fig. S3c) and the corresponding angular tilt in the longitudinal direction (Fig. S3d).

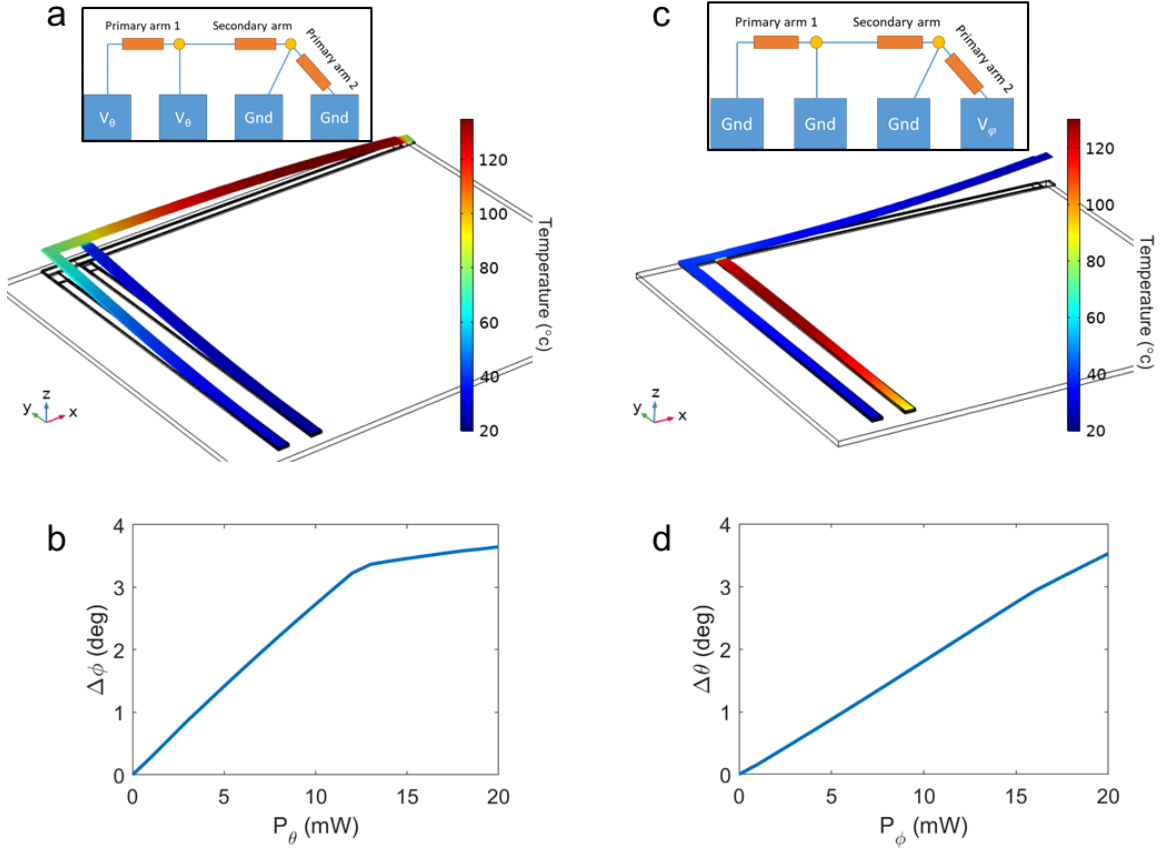

**Fig. S3: Electro-thermo-mechanical simulation of the L-shaped cantilever. a** Simulated temperature overlaid on the displaced L-shaped cantilever when electrical power of $P_\theta$ = 10 mW is applied to the secondary arm. **b** Simulated angular tilt of the grating coupler in transverse direction under the electrical power applied to the secondary arm. The insets in **a** and **b** show the circuit diagrams. **c** Simulated temperature of the L-shaped cantilever overlaid on its displacement under electrical power of $P_\varphi$ = 10 mW applied to the right-hand side primary arm. **d** Calculated φ-axis angular tilt of the grating coupler as a result of the voltage applied to the right primary arm.

### 4. Time response measurements

Using the procedure described in the Methods Section of the main manuscript, we measured the temporal response of several cantilevers. Fig. S4a shows a few recorded far-field trajectories of the 800 μm long rectilinear cantilevers, taken with different duty cycles. In the measurements in Fig. S4a, the time period of the signal was set to 20 ms. The measured and simulated time responses of the shortest rectilinear cantilever are shown in the main manuscript (Fig. 3b). Figures S4b, c, d show the extracted rise time and fall times of the 500 μm, 800 μm, and 1 mm long rectilinear cantilevers. In these measurements, the applied square pulse had a period of 20 ms and the duty-cycle was varied between 1% to 50%. Since the temporal responses of the devices were measured at discrete points (blue dots in Fig. S4), we fit an exponential curve to the measured values (red dashed lines). The rise time (/fall time) of the cantilevers were measured to be 1.01 (/1.41), 2.42 (/2.84), 4.14 (/4.03), and 3.63 (/5.72) ms respectively for 300, 500, 800, and 1000 µm long cantilevers, in good agreement with the simulated (yellow dashed lines in Fig. S4) time response.

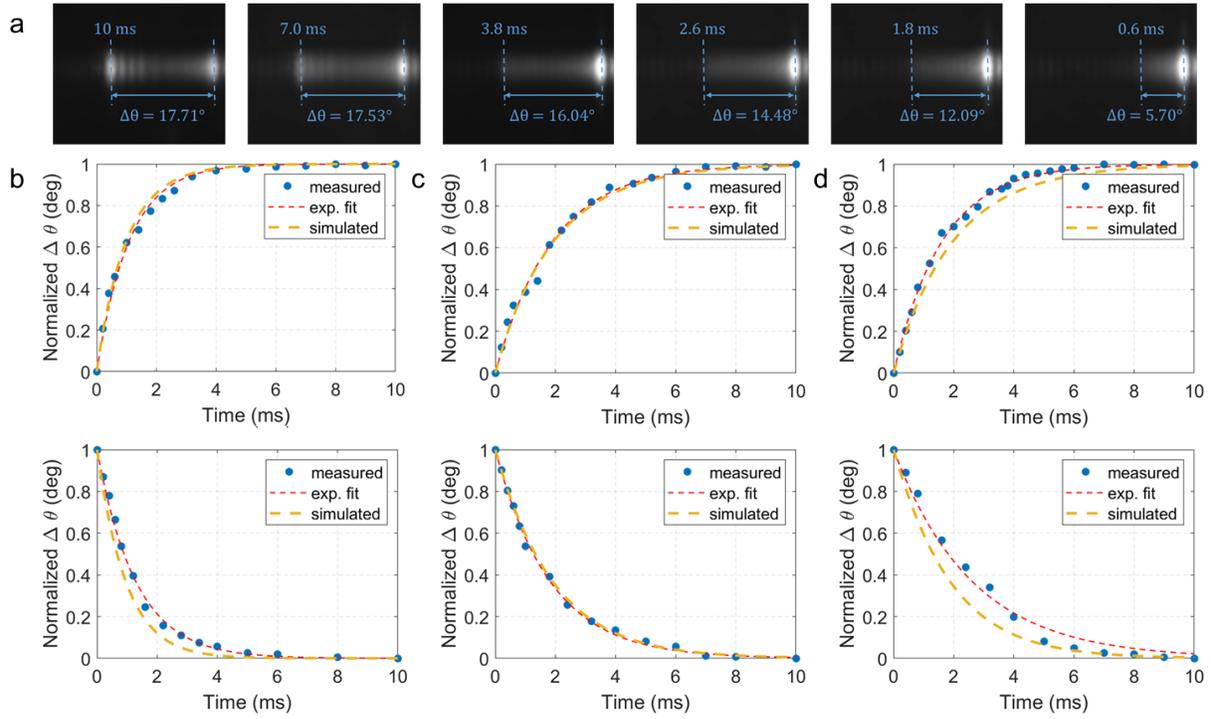

**Fig. S4: Time response of the rectilinear cantilevers. a** Recorded far-field images of the output of the 800 µm long rectilinear cantilever, under applied square signal with maximum power of 10 mW, period of 20 ms, and duty cycles of 50%, 35%, 19%, 13%, 9% and 3%, respectively, from left to right. **b**, **c, d** Measured and simulated rise time (top) and fall time (bottom) of the 500 µm (**b**), 800 µm (**c**), and 1 mm (**d**) long cantilevers.

We also measured the time responses of the L-shaped cantilevers. The results are shown in Figs. S5. In this case, we separately measured the temporal response of the secondary arm (Fig. 5Sa) and the right-hand side primary arm (Fig. S5b). We measure the rise time of 4.26 and 4.65 ms and fall time of 4.87 and 5.64 ms respectively for the primary and the secondary arms.

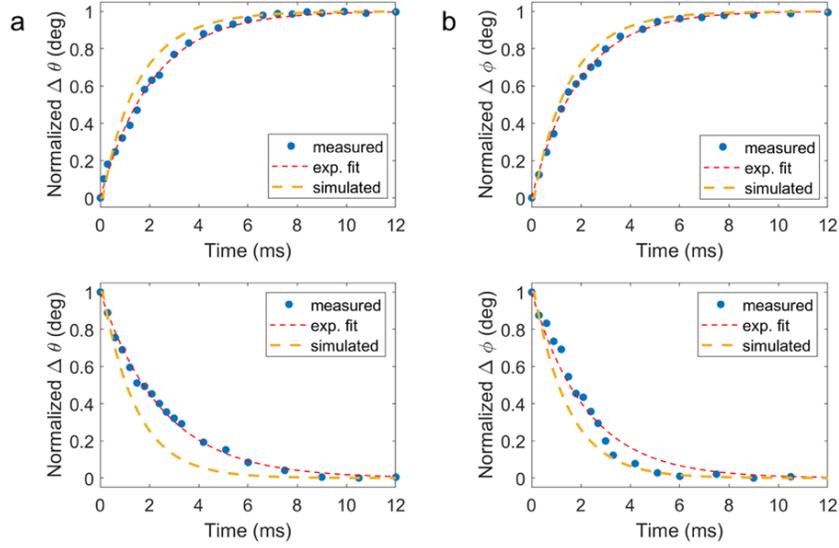

**Fig. S5: Time response of the L-shaped cantilever.** Measured and simulated rise time (top) and fall time (bottom) of the secondary arm (**a**) and primary arm (**b**) of the L-shaped cantilever.

## 5. Improving the resolution by modifying the grating coupler design

The number of resolvable points of our devices was limited due to the short length (25 µm) of the grating couplers, which led to a relatively large divergence angle (FWHM = 1.4°). The number of resolvable points, at least in the longitudinal direction, could be significantly increased without modification of the cantilever design by increasing the grating coupler length along the cantilever and reducing the scattering strength of the grating. To quantitatively illustrate the effectiveness of this approach, based on the analytical model explained in Chapter 6 of [10], we calculated the FWHM of the output beam at λ=488 nm as a function of grating coupler length ($L_{GC}$) and the scattering strength (α) of the gratings. The calculation results of the analytical model are shown in Fig. S6a for $L_{GC}$ between 5 µm and 2 mm. For large values of the scattering strength (e.g., blue curve in Fig. S6a), increasing the grating length beyond the point where the light is fully coupled out does not decrease the FWHM of the output beam. This is the case for single layer fully-etched grating couplers, which were used in this work. However, by using sidewall gratings, it is possible to effectively increase the grating length and reduce the FWHM. To validate the analytical method, we simulated a 100 µm long and 4 µm wide SiN grating coupler with additional 2 µm wide sidewall gratings, using the same thickness of 150 nm, using 3D FDTD simulation in Lumerical. The resulting output beam in the propagation direction is shown in Fig. S6b corresponding to the green dot in Fig. S6a. Compared to the grating couplers we used (red dot in Fig. 6Sa) the simulated grating coupler has a 4-fold improvement in the FWHM in the longitudinal dimension. The design flexibility of the demonstrated cantilevers allows for tuning the properties of the output beam for a certain application. For instance, since a high angular resolution is desirable for head-mounted augmented reality displays, the length of the grating can be increased to 1 mm (see Fig. S6a) to each a beam size of 0.02°.

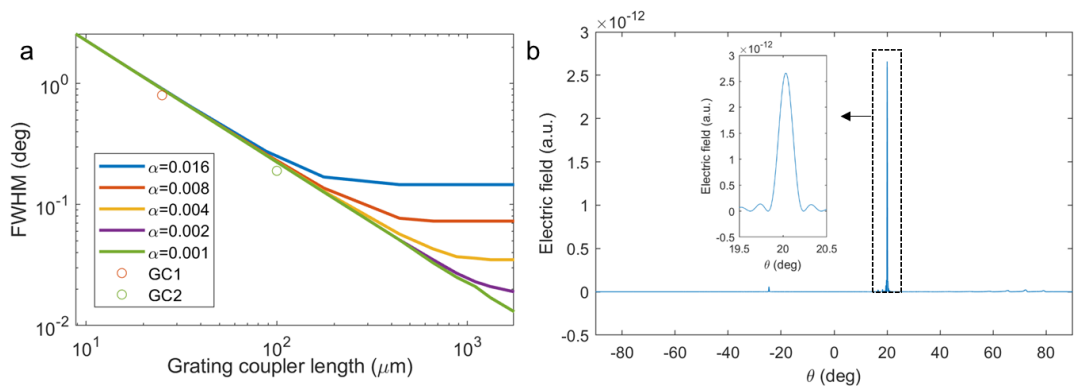

**Fig. S6: Simulated performance of a weaker grating coupler. a** Calculated full-width-half-maximum of the output beam of the grating coupler according to its length and scattering strength. The red and green dots, respectively, represent the measured 25 µm long grating coupler (GC1) and the simulated and the simulated 100 µm long grating coupler (GC1) **b** Simulated far-field profile of the output beam at λ=488 nm for a 100 µm long grating coupler with α=0.002.

## 6. Cryogenic measurement setup of the L-shaped microcantilever

The characterization setup used for cryogenic measurements is shown in Fig. S7a. The cryostat had an open window of 10 cm in diameter at the top that allowed us to view and measure the emitted beam. The optical and electrical packaging of the PIC were performed prior to the measurements as explained in the Methods section of the main manuscript. Generally, electro-thermally tuned optical devices are not designed to perform in cryogenic conditions. For instance, the high thermal budget of the state-of-the-art SiN beam formers [11] can easily overwhelm the cooling power of the cryostat which is typically in the range of hundreds of milliwatts. However, due to the relatively small power consumption of our devices, the temperature could be reduced to 4K for an applied power of < 30 mW. We performed thermal simulations using COMSOL Multiphysics to calculate the temperature profile in the PIC under 26 mW of applied electrical power to the secondary arm of the L-shaped cantilever, while the chip is fixed on the cold head with T = 10 K. The results (Figs. S7b and S7c) show that at a distance of only ~ 600 µm from the cantilever, the temperature falls to 11 K, which is one degree above the set point of the cryostat. This suggests that our devices can be implemented on the same chip in conjunction with other electro-optical devices (such as single-photon detectors) which require a low-temperature condition for their performance, without any significant crosstalk, provided that an adequate clearance is maintained.

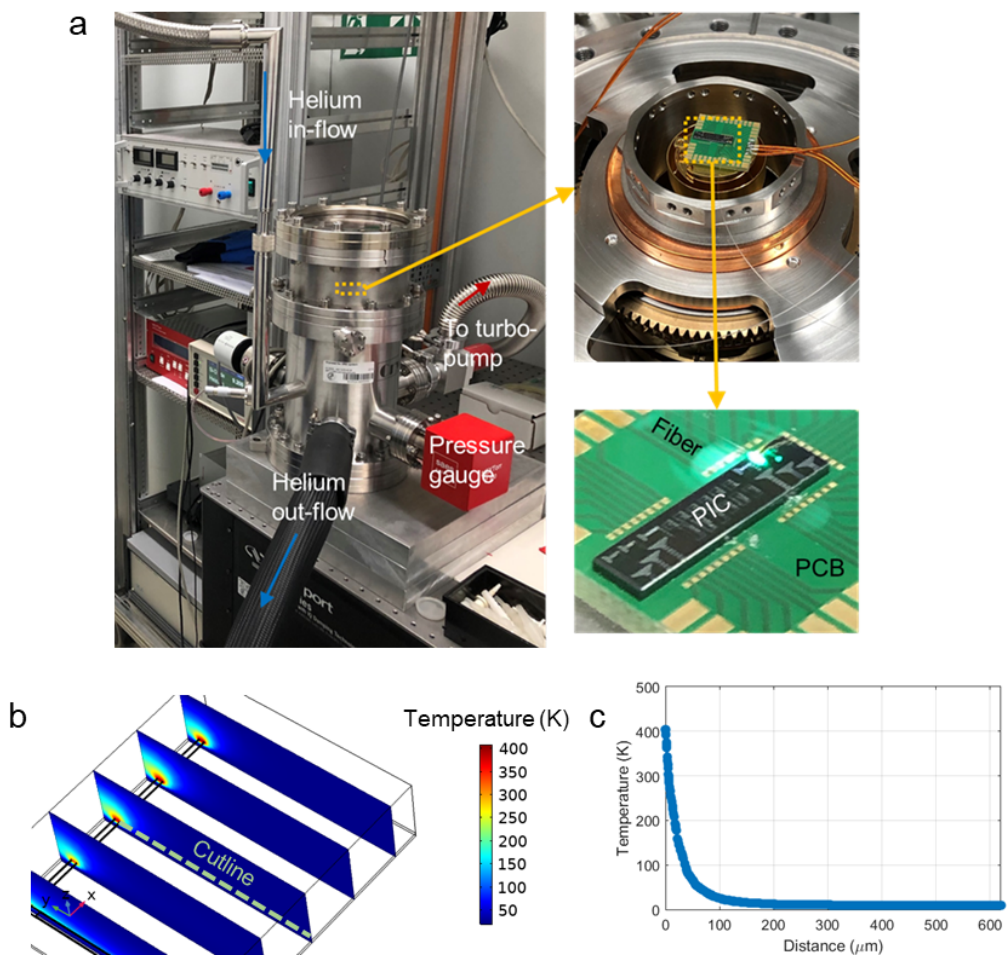

**Fig. S7: a** Cryogenic measurement setup. **b** Simulated temperature profile in the vicinity of the L-shaped beam scanner under 26 mW of applied electrical power. **c** Temperature along the green dashed line in **b**) showing the decay of the temperature.

## 7. Comparison with state-of-the-art beam steering systems

The results of the rectilinear and L-shaped cantilevers in terms of steering range, power consumption, number of resolvable spots, temporal response, and resonant frequencies are summarized in Table S1.

**Table S1 | Summary of the demonstrated devices**

| Device | Dimension | Steering range (°) | Resolvable points | Required power (mW) | $t_{rise}$ (ms) $t_{fall}$ (ms) | $f_{res}$ (kHz) |
|---|---|---|---|---|---|---|
| L-shaped (500 × 600 μm) | 2D | 24 × 12 | 66 | 45 | 4.3, 4.7 4.9, 5.6 | 7.6 ($f_1$) 17.4 ($f_2$) |
| Rectilinear (300 μm) | 1D | 11 | 8 | 30 | 1.0 1.4 | 77.4 |
| Rectilinear (500 μm) | 1D | 17.6 | 12 | 30 | 2.4 2.8 | 24.8 |
| Rectilinear (800 μm) | 1D | 22.6 | 16 | 30 | 4.1 4.0 | 11.8 |
| Rectilinear (1 mm) | 1D | 30.1 | 21 | 30 | 3.6 5.7 | 5.7 |

Table S2 presents a comparison of this work with other beam-steering PIC demonstrations in the visible spectral range. So far, only a few PIC beam scanners have been reported. The microcantilevers presented here achieve fast beam steering without a wavelength sweep and at a record low power consumption.

**Table S2 | Comparison of beam steerers in the visible spectrum**

| Device | Dim. | λ(nm) | Beam steering method | Steering range (°) | Resolvable points | Required power (mW) | Time response | $f_{res}$ (kHz) | Emitter loss | Ref. |
|---|---|---|---|---|---|---|---|---|---|---|
| L-shaped cantilever | 2D | 410 -700[1] | θ: MEMS<br>φ: MEMS | 24 × 12 | 66 | θ: 23<br>φ: 23 | ~5 ms | 7.6 ($f_1$)<br>17.4 ($f_2$) | 5.2 dB | This work |
| Optical phased array | 1D | 520 -980 | θ: Wavelength sweep | 65 | ~260[2] | Not reported | Not reported | NA | Not reported | [12] |
| Optical phased array | 1D | 488 | φ: Thermal tuners | 50 | ~294[2] | φ: 2000 | Not reported | NA | Not reported | [11] |
| Optical phased array | 2D | 650 -980 | θ: Wavelength sweep<br>φ: Thermal tuners | 44 × 13 | Not reported[3] | θ: Not rep.<br>φ: Not rep. | Not reported | NA | Not reported | [13] |

[1] Two-dimensional beam steering can be achieved at any target wavelength in this range.
[2] Calculated based on the reported FWHM.
[3] FWHM in the longitudinal direction is reported to be 3.4° at λ = 850 nm.